\documentclass[tradiabstract]{aa}
\usepackage{graphicx}
\usepackage{txfonts}
\usepackage{natbib}
\usepackage{journal_shortcuts}
\newcommand{\be}{\begin{equation}}
\newcommand{\ee}{\end{equation}}
\newcommand{\bea}{\begin{eqnarray}}
\newcommand{\eea}{\end{eqnarray}}
\newcommand{\CW}{c_{\mathrm{w}}}

\newcommand{\ICM}{_{\mathrm{ICM}}}

\newcommand{\Amb}{_{\mathrm{amb}}}
\newcommand{\Eff}{_{\mathrm{eff}}}
\newcommand{\Gas}{_{\mathrm{gas}}}

\newcommand{\Cloud}{_{\mathrm{cl}}}
\newcommand{\Ram}{_{\mathrm{ram}}}

\newcommand{\Kpc}{\,\textrm{kpc}}

\newcommand{\CM}{\,\textrm{cm}}

\newcommand{\Kms}{\,\textrm{km}\,\textrm{s}^{-1}}

\newcommand{\ccm}{\,\textrm{cm}^{-3}}
\newcommand{\gccm}{\,\textrm{g}\,\textrm{cm}^{-3}}

\begin{document}
\title
{The role of the Rayleigh-Taylor instability in ram pressure stripped 
disk galaxies}

\titlerunning{RT instability in ram pressure stripped disk galaxies}

\author{E. Roediger\inst{1,2}
       \and
       G. Hensler\inst{3}
}

\authorrunning{Roediger \& Hensler}

\institute{Jacobs University Bremen, P. O. Box 750 561, 28725 Bremen,
           Germany\\
    \email{e.roediger@jacobs-university.de}
    \and
       Institute of Theoretical Physics and Astrophysics,
           University of Kiel,
           Olshausenstr. 40, D-24098 Kiel, Germany
    \and %
          Institute of Astronomy, University of Vienna,
          T\"urkenschanzstrasse 17, A-1180 Vienna, Austria\\
        \email{hensler@astro.univie.ac.at}
}

\date{Received; accepted}

\abstract {Ram pressure stripping, i.e. the removal of a galaxy's gas
  disk due to its motion through the intracluster medium of a galaxy cluster,
  appears to be a common phenomenon. Not every galaxy, however, is completely
  stripped of its gas disk. If the ram pressure is insufficiently
  strong, only the outer parts of the gas disk are removed, and the inner gas
  disk is retained by the galaxy. One example of such a case is the Virgo
  spiral NGC~4402. Observations of NGC~4402 (\citealt{crowl05}) reveal
  structures at the leading edge of the gas disk, which resemble the
  characteristic finger-like structures produced by the Rayleigh-Taylor (RT)
  instability. We argue, however, that the RT instability is unlikely to be
  responsible for these structures. We demonstrate that the conditions under
  which a galaxy's disk gas experiences ram pressure stripping are identical
  to those that lead to RT instability. If the galaxy's gravity prevents ram
  pressure stripping of the inner disk, it also prevents the RT
  instability. In contrast, the stripped gas could still be subject to RT
  instability, and we discuss consequences for the stripped gas.


}
\maketitle

%
%
\section{Introduction}
%
Differences between field and cluster galaxies imply that the evolution of
galaxies is influenced strongly by their environments. In addition to the
gravitational interaction between the cluster galaxies themselves, ram
pressure stripping (RPS), the (partial) removal of a galaxy's gas due to its
motion through the intracluster medium (ICM), is thought to play an important
role for the evolution of galaxies (see e.g.~\citealt{vangorkom04}).

RPS has been modelled numerically by different groups
(e.g.~\citealt{abadi99,schulz01,quilis00,vollmer01a,marcolini03};
\citealt{roediger05}, hereafter RH05; \citealt{roediger06}, hereafter RB06;
\citealt{roediger06wakes}; \citealt{roediger07}, hereafter RB07). In
environments in which ram pressure stripping is prevalent, such as cluster
centres, the stripping is very effective, but the central part of the gas disk
can be retained by the galaxy for quite some time.

In principle, the remaining gas disk can be affected by the Rayleigh-Taylor
(RT) instability, which occurs when a lighter fluid supports a heavier one
against a gravitational field or an acceleration. An equivalent situation
occurs when a denser fluid is accelerated by a lighter one, e.g.~when a dense
gas cloud is exposed to a wind of rarefied gas (or when a gas cloud moves
through this ambient gas). Such a gas cloud is expected to be destroyed by the
RT instability, and subsequent Kelvin-Helmholtz (KH) instability, unless
stabilising mechanisms are in place such as (self-)gravity, surface tension,
magnetic fields (\citealt{chandrasekhar61}), or heat conduction
(\citealt{vieser07}).

Thus, as a (disk) galaxy moves through the ICM and its gas is
stripped by RPS, it should be affected by the RT instability. The detection of
a filamentary leading surface in \object{NGC 4402}, a Virgo cluster galaxy with clear
signs of RPS, lends support to the idea that the RT instability does affect
the remaining gas disk. Recent images by \citet{crowl05}
reveal finger-like structures at the galaxy's upstream side, which are
slightly inclined with respect to the rotation axis of the disk.
 If these pillars are interpreted as the results of the RT
instability, the direction of motion of the galaxy that they imply would be
identical to that implied by HI observations (\citealt{crowl05}).

Such structures are not, however, seen in the simulations.  In
Sect.~\ref{sec:remaining-disk}, we demonstrate analytically that the RT
instability is suppressed in the remaining gas disk due to the gravitation of
the galaxy. In addition, we discuss the role of the RT instability in deciding
the fate of stripped gas in Sect.~\ref{sec:stripped}. To start with, we
summarise basic analytical estimates concerning the RT instability in
Sect.~\ref{sec:rt}.

\section{The Rayleigh-Taylor instability}
\label{sec:rt}
A detailed derivation of stability conditions and instability growth rates is
provided by \cite{chandrasekhar61}.

We consider the simplest case of two superimposed, incompressible, inviscid,
unmagnetised fluids. These fluids are separated by a contact discontinuity
that is perpendicular to the effective homogeneous (gravitational)
acceleration, $a$. Referring to the direction of $a$ as ``downwards'', this
setup is unstable if the upper fluid has a higher density than the lower
one. Any perturbation at the interface will be amplified by the RT
instability, developing characteristic finger-like structures. In the linear
regime, small perturbations grow at the most rapid rate. Also a continuous
stratification will be unstable if the density increases upwards.

\begin{figure}
\resizebox{\hsize}{!}{
\includegraphics[width=\textwidth,angle=0]{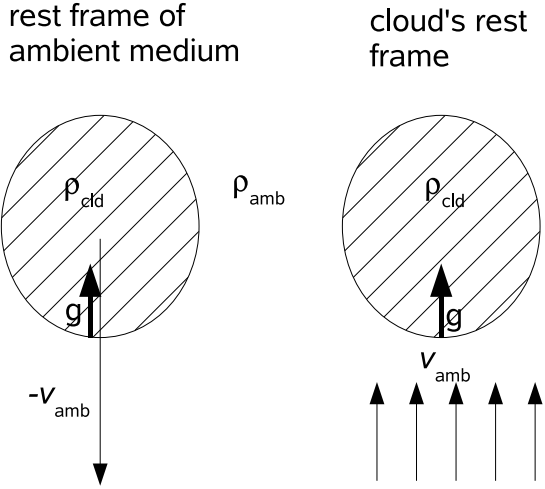}}
\caption{Sketch of a cloud moving through an ambient medium, in the rest frame
  of the ambient medium (left), and in the cloud's rest frame (right). The
  direction of the gravitational acceleration at the cloud's upstream side, due
to its self-gravity, is shown by the vector $g$.}
\label{fig:sketch}
\end{figure}
%
Here we are interested in the situation sketched in Fig.~\ref{fig:sketch}: a
dense cloud of mass, $M\Cloud$, density, $\rho\Cloud$, and cross section $A$
is moving with velocity, $-v\Amb$, through an ambient
gas of density $\rho\Amb$, or, equivalently, is exposed to a wind of ambient
gas with velocity $v\Amb$. The relevant acceleration
is the one exerted by the drag force $F_D$:
\bea 
a_D&=&\frac{F_D}{M\Cloud} = \frac{ \frac{1}{2}c_W A \rho\Amb v\Amb^2}{M\Cloud}
\nonumber \\
&=& \frac{ \frac{1}{2}c_W \rho\Amb v\Amb^2}{\Sigma\Cloud}
=\frac{1}{2}c_W \frac{p\Ram}{\Sigma\Cloud} \; ,
\eea
where $c_W$ is the drag coefficient, and the mean
column density of the cloud is given by
\be
\Sigma\Cloud =\frac{M\Cloud}{A}=\frac{4/3 \pi \, R\Cloud^3 \, \rho\Cloud}{\pi
  R\Cloud^2}=4/3 \, R\Cloud \, \rho\Cloud.
\label{eq:sig}
\ee
The drag coefficient depends on the cloud's shape
and surface properties, where $\frac{1}{2}c_W \sim 1$ may be used as an
approximation. At the upstream side of the cloud, in the rest frame of the
cloud the drag force appears as a pseudo force pointing upstream, the correct
direction to produce the RT instability at the cloud's upstream
edge.

A possible way to stabilise the cloud's surface against the RT instability
would be a gravitational field associated with the cloud. The gravitational
acceleration, $g$, at the cloud's upstream edge works in the opposite
direction of $a_D$ (``inwards'' for the cloud), so that the relevant
acceleration for the development of the RT instability is the effective
acceleration
%
\be 
a\Eff= -a_D + g  \; .
\ee
%
The RT instability will be suppressed, if $a\Eff > 0$, or
%
\be 
g > a_D \; . \label{eq:grav_suppr}
\ee

Another mechanism to stabilise the cloud against RT instability, is provided
by magnetic fields perpendicular to the cloud surface. Magnetic fields act
like a surface tension and suppress the growth of perturbations of
scale-length $\lambda < \lambda_c$ with
%
\be
\lambda_c = \frac{B^2\; \cos^2\theta}{\mu_0 (\rho\Cloud - \rho\Amb) a_D}
=\frac{B^2\; \cos^2\theta \;\Sigma\Cloud}{\mu_0 \rho\Cloud \frac{1}{2}c_W
  \rho\Amb v\Amb^2} \; ,    \label{eq:magn_suppr}
\ee
%
where $\theta$ is the angle between the magnetic field direction and the
direction of the perturbation (\citealt{chandrasekhar61}). We assume that
$\rho\Cloud \gg \rho\Amb$.  If $\lambda_c$ is larger than the size of the
cloud, it should therefore be resistant to RT instability because the largest
possible perturbations have a size that is similar to that of the cloud.

\section{Application}
%
\subsection{The remaining gas disk} 
\label{sec:remaining-disk}
We now investigate to what extent the RT instability can develop in a disk
galaxy that is either moving face-on through the ICM, or is exposed to an ICM
wind of density $\rho\ICM$ and velocity $v\ICM$. In Fig.~\ref{fig:sketch}, we
replace the cloud by a disk galaxy with gas surface density, $\Sigma\Gas$.
The gravitational acceleration due to the galaxy potential, $g$, at the
upstream side of the galaxy, points downstream. According to
Eq.~\ref{eq:grav_suppr}, the condition for RT instability to {\em occur} is
%
\bea 
g &<& \frac{1}{2} \CW \frac{\rho\ICM \,  v\ICM^2}{\Sigma\Gas}\\
\textrm{or}\qquad g\Sigma\Gas &<& \frac{1}{2} \CW \rho\ICM \, v\ICM^2 \; . 
\label{eq:RT}
\eea
%
For a disk galaxy, both, $g$ and $\Sigma\Gas$ depend on the radial position
within the disk and both typically decrease with disk radius, $r$.  In the
outer parts (where $g\Sigma\Gas$ is smaller), the RT instability can occur,
while it is suppressed in the inner part, where $g\Sigma\Gas$ is larger.  An
identical criterion is applied to decide the radius out to which gas will be
stripped from the disk by ram-pressure (see e.g.~RH05, RB06). The ram
pressure, $\frac{1}{2} \CW \rho\ICM \, v\ICM^2$, can be compared to the
gravitational restoring force, $g\Sigma\Gas$, as written in Eq.~\ref{eq:RT}.
At radii where this inequality applies, the gas will be ram pressure stripped;
if the inequality does not apply, the expression on its left-hand side will
dominate, and the galaxy retains its gas. In this second case, the inner gas
disk is not stripped and is protected against RT instability by the gravity of
the galaxy.  This behaviour is reproduced by simulations (see
e.g.~Fig.~\ref{fig:fragmentation}).

We note that tangential magnetic fields would stabilise the remaining disk
even further.

\subsection{The stripped gas} 
\label{sec:stripped}
%
\begin{figure}
\resizebox{\hsize}{!}{
\includegraphics[width=0.5\textwidth,angle=0]{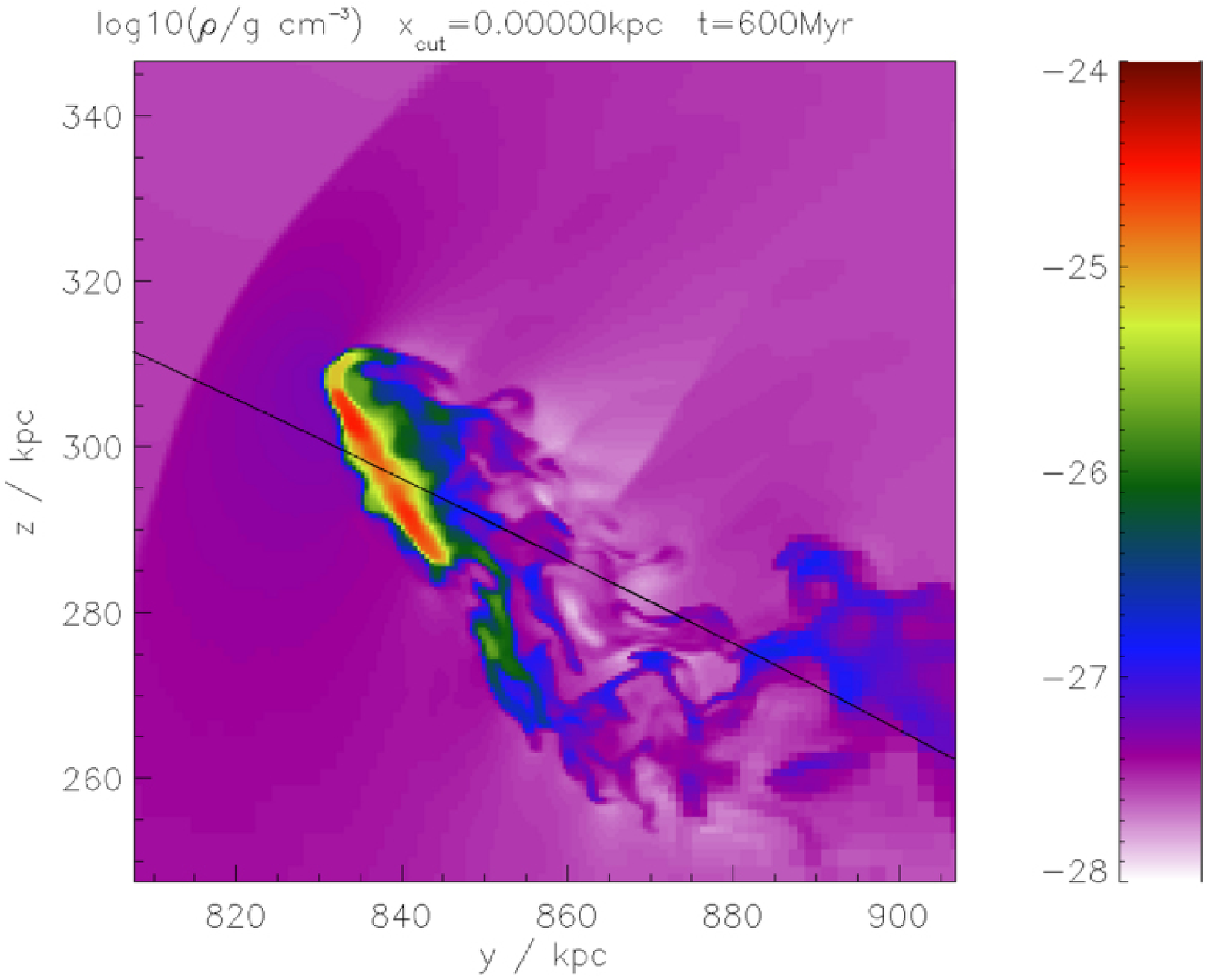}
\includegraphics[width=0.5\textwidth,angle=0]{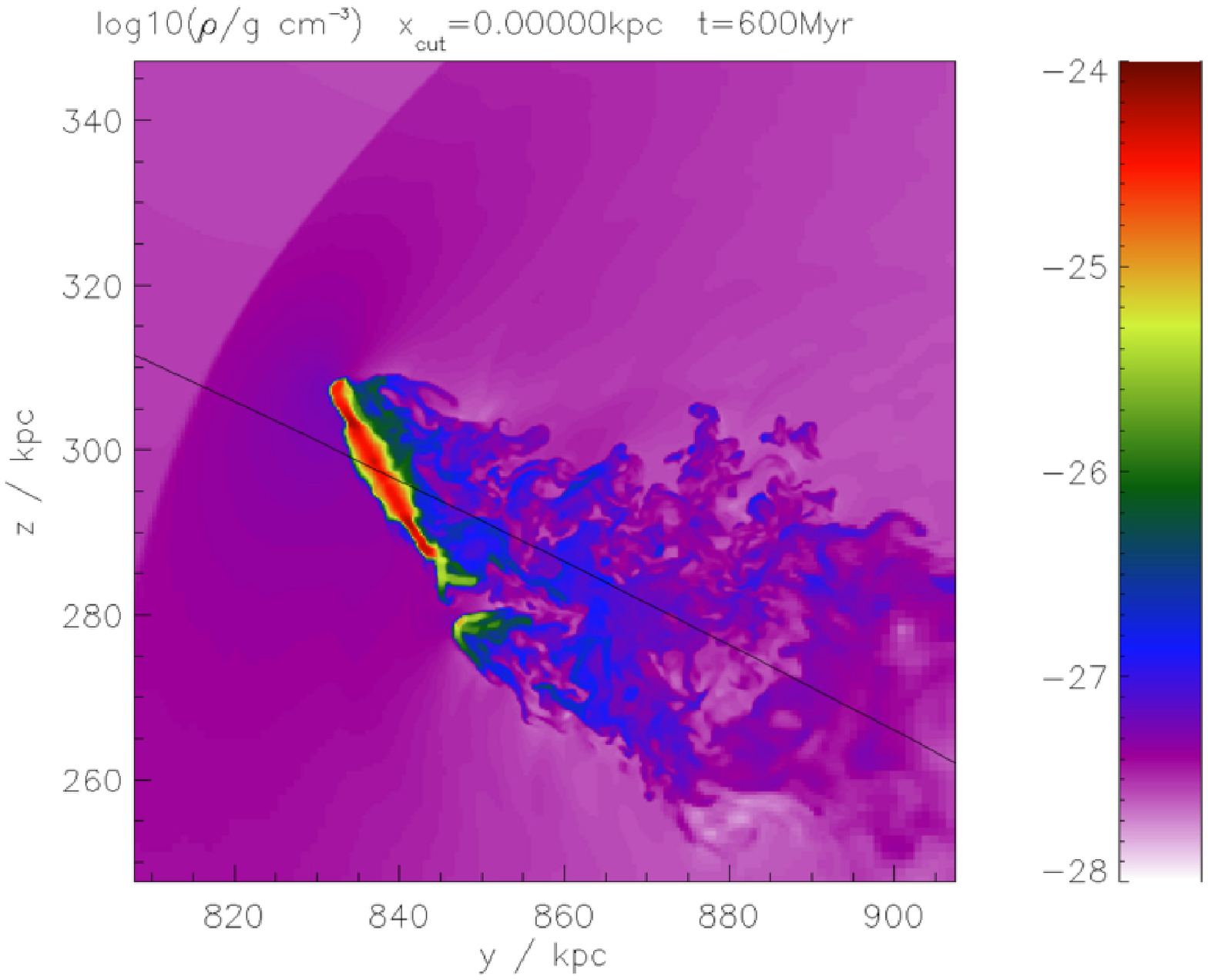}}
\caption{Simulation snapshots from simulations described in RB07. Identical
 conditions except for the resolution: in rhs panel, the spatial resolution is 2
 times better than in the lhs panel. The remaining gas disk is not fragmented,
 but the stripped gas is. The fragmentation is stronger for higher
 resolution.}
\label{fig:fragmentation}
\end{figure}
%
According to the results of the previous section, the galaxy's gravitational
field cannot prevent the RT instability of the stripped material. In the
absence of other stabilising effects, the stripped gas should therefore
fragmentate.  The simulations of e.g.~RH05, RB06 and RB07 neglected all
possible stabilising effects, and therefore led to
fragmentation. Figure~\ref{fig:fragmentation} displays two simulation
snapshots that differ only in spatial resolution.  The resolution fixes the
lengthscale of the smallest possible perturbations.  According to analytical
estimates, the smallest perturbations grow most rapidly, thus the stripped gas
should fragmentate into smaller clouds, in simulations of higher
resolution. This behaviour is in fact observed in, both, our 2D and 3D
simulations.

\subsubsection{Self-gravity} 
\label{sec:application-selfgrav}

We consider the extent to which self-gravity of stripped gas clouds can
prevent RT instability. This case will apply at the cloud's surface, if the
gravitational acceleration exceeds $a_D$ (Eq.~\ref{eq:grav_suppr}). If we
assume a spherical gas cloud of density, $\rho\Cloud$, and radius, $R\Cloud$,
after completing some simple arithmetic, we find that the RT instability is
suppressed if
%
\bea
&&\frac{\Sigma\Cloud}{3\cdot 10^{21}\CM^{-2}}=\frac{n\Cloud}{\ccm} \frac{R\Cloud}{\Kpc} > \nonumber\\
&& \frac{1}{2} \; 
\left({\frac{\rho\ICM}{10^{-27}\gccm}}\right)^{1/2}
\left({\frac{v\ICM}{1000\Kms}}\right) \left(\frac{c_W}{0.5}\right)^{1/2}
\eea
%
This implies that only clouds of high column densities can prevent an RT
instability. We remark, however, that these clouds may yet be affected and
dissolved, by the KH instability.

\subsubsection{Magnetic fields} \label{sec:application-magnetic}

A second mechanism to stabilise stripped-off clouds against RT instability are
tangential magnetic fields. We assume again a spherical gas cloud of radius,
$R\Cloud$, and substitute the expression for $\Sigma\Cloud$, from
Eq.~\ref{eq:sig} into Eq.~\ref{eq:magn_suppr} to obtain
%
\bea
\lambda_c &=& 
\frac{B^2\; \cos^2\theta\; \frac{4}{3} R\Cloud} 
{\mu_0\; \frac{1}{2}c_W\; \rho\ICM\; v\ICM^2} 
=  1.06 \cdot R\Cloud \; \cos^2\theta  \left(\frac{B}{5\mu G}\right)^2 \times\nonumber\\
&&\left(\frac{\rho\ICM}{10^{-27}\gccm}\right)^{-1} 
\left(\frac{v\ICM}{1000\Kms}\right)^{-2} \left(\frac{c_W}{0.5}\right)^{-1} \; .
\label{eq:lambdac}
\eea
%
This result is independent of the density of the cloud, and compares the
energy density of the ICM wind with the energy density of the magnetic
field. The magnetic field in spiral galaxies consists of an ordered
large-scale field of approximately $1\mu G$, and a small-scale, tangled
component with a field strength of about $5\mu G$ (see \citealt{lequeux05} and
references therein). We therefore assume that because of the small-scale
field, a magnetic field $B$-component tangential to the cloud's surface with
field strength of the order of a few $\mu G$ will exist. The critical value of
$\lambda_c$ is then of the same order as the radius of the cloud. It is
therefore likely that the stripped gas will suffer RT instability only in case
of strong ICM winds.

\subsubsection{Kelvin-Helmholtz-instability} 
\label{sec:application-KH}

In addition to the RT instability, the stripped gas can suffer KH
instability. This instability cannot be prevented by a gravitational field,
but can be averted by tangential magnetic fields. We note that the relevant
surface for KH instability are the sides of the gas clouds. For
tangled magnetic fields however, a tangential component should be found. The
criterion to suppress the KH instability is similar to before. It is
suppressed when
%
\bea
1&<& \frac{B^2}{2 \pi\; \mu_0\; \rho\ICM\; v\ICM^2} 
= 0.037 \left(\frac{B}{5\mu G}\right)^2 \times\nonumber\\
&&\left(\frac{\rho\ICM}{10^{-27}\gccm}\right)^{-1} 
\left(\frac{v\ICM}{1000\Kms}\right)^{-2} \; ,
\label{eq:B}
\eea
%
implying that the KH instability occurs in weaker winds, even if the
RT instability does not take effect, apart from very weak winds
such as those in the cluster outskirts.

\section{Discussion}
%
In this paper, we have derived two main results.  We have shown that the inner
parts of ram pressure stripped disk galaxies, which are resistant to RPS, are
in addition stable against RT instability in the case of face-on
stripping. The same should be true at mild inclinations, as neither the
gravitational acceleration in the wind direction, nor the projected gas
surface-density changes significantly with inclination angle.  In addition,
the clumps of stripped-off gas are exposed to RT instability and should be
destroyed.

Given that the RT instability is suppressed for the inner gas disk that
survives ram pressure stripping, we conclude that the structures observed at
the leading edge of NGC~4402 are not caused by the RT instability.  These
structures could be caused by different processes, one possibility being the
inhomogeneous ISM as discussed by Crowl et al. (2005). Quilis et al. (2000)
first studied the effect of an inhomogeneous ISM on ram-pressure stripping
simulations, by creating holes inside the gas disk.  The holes produced more
pronounced Kelvin-Helmholtz instabilities, and thus stronger mass loss. The
observations of NGC~4402 do not concern under-dense but over-dense regions in
the ISM: dense gas clumps appear to be left behind after the diffuse gas is
stripped, but the dense clouds also appear to be ablated. To follow such
structures in simulations requires extremely high resolution and computational
effort. Nonetheless, this is an important task for future simulations.

To what extent could stripped gas be observed? A complete investigation of
this question would have to include processes such as thermal conduction and
evaporation, cooling and star formation, as well as the effects of RT and KH
instabilities.  The above estimates imply that the stripped gas is likely
affected by RT instability, and possibly also KH instability.  Nevertheless,
more massive stripped-off gas packages can be gravitationally bound and stable
against RT instability. Qualitatively, this can explain why only a few massive
clouds are visible (in H$_\alpha$) for galaxies observed being affected by RPS
(such as e.g.~NGC~4522: \citealt{kenney99}: and NGC~4569: Fig.\ 4 in
\citealt{tschoeke01}) while in smaller clumps the destructive RT instability
dominates and disperses the gas into the ICM.  The fact that the stable,
stripped-off clouds emit H$_\alpha$ in spite of no stars having formed,
implies that they are heated by external sources.  As the most plausible
candidate, heat conduction should be suggested because, in contrast to general
assumptions, self-gravitating clouds are not destroyed by evaporation but
accumulate gas by condensation (\citealt{vieser07}), and are, furthermore,
stabilised against KH instability.  \citet{vollmer01a} discussed the influence
of thermal conduction and effect of reduced heating on stripped gas
clouds. They argued that the clouds no longer absorb the stellar
far-ultraviolet radiation field. They conclude that the stripped gas could be
either ionised and hot, or in the form of cool and even molecular clouds,
enclosed in neutral atomic shells.  More detailed studies are required to
study the combined influence of these effects, as well as the influence of
star formation.

\section*{Acknowledgements}
This work was supported by the \emph{Deut\-sche
For\-schungs\-ge\-mein\-schaft\/} project number He~1487/30. We
gratefully acknowledge fruitful and helpful discussions with Marcus Br\"uggen,
Joachim K\"oppen, Bernd Vollmer and Curtis Struck.

%

\bibliographystyle{aa}
\bibliography{%
../../../BIBLIOGRAPHY/theory_simulations,%
../../../BIBLIOGRAPHY/hydro_processes,%
../../../BIBLIOGRAPHY/numerics,%
../../../BIBLIOGRAPHY/observations_general,%
../../../BIBLIOGRAPHY/observations_clusters,%
../../../BIBLIOGRAPHY/observations_galaxies,%
../../../BIBLIOGRAPHY/galaxy_model,%
../../../BIBLIOGRAPHY/gas_halo,%
../../../BIBLIOGRAPHY/icm_conditions,%
../../../BIBLIOGRAPHY/ism,%
../../../BIBLIOGRAPHY/else}

\end{document}